\RequirePackage[2020-02-02]{latexrelease}
\documentclass[aip,amsmath,amssymb,reprint]{revtex4-1}
\usepackage{graphicx}
\usepackage{dcolumn}
\usepackage{xcolor}
\usepackage{bm}
\usepackage{enumerate}
\usepackage{threeparttable}
\usepackage{multirow}
\usepackage[normalem]{ulem}
\usepackage[utf8]{inputenc}
\usepackage[T1]{fontenc}
\usepackage{mathptmx}
\usepackage{braket}
\usepackage{xr}
\usepackage{booktabs}
\usepackage{tabu}
\usepackage{float}
\usepackage{interval}
\usepackage{natbib}
\makeatletter

\def\dd{{\rm d}}

\newcommand*{\addFileDependency}[1]{
  \typeout{(#1)}
  \@addtofilelist{#1}
  \IfFileExists{#1}{}{\typeout{No file #1.}}
}
\makeatother

\newcommand*{\myexternaldocument}[1]{
    \externaldocument{#1}
    \addFileDependency{#1.tex}
    \addFileDependency{#1.aux}
}

\myexternaldocument{SI}
\begin{document}

\preprint{AIP/123-QED}

\title{Embedding vertex corrections in $GW$ self-energy: theory, implementation, and outlook}

\author{Guorong Weng}
\author{Rushil Mallarapu}
\altaffiliation[Current institution: ]{Harvard University, Cambridge, MA, 02138, USA } 
\author{Vojt\v{e}ch Vl\v{c}ek}
\email{vlcek@ucsb.edu}
\affiliation{Department of Chemistry and Biochemistry, University of California, Santa Barbara, CA 93106-9510, USA}

\date{\today}

\begin{abstract}
The vertex function ($\Gamma$) within the Green's function formalism encapsulates information about all higher-order electron-electron interaction beyond those mediated by density fluctuations. Herein, we present an efficient approach that embeds vertex corrections in the one-shot $GW$ correlation self-energy for isolated and periodic systems. The vertex-corrected self-energy is constructed through the proposed separation-propagation-recombination procedure: the electronic Hilbert space is separated into an active space and its orthogonal complement denoted as the ``rest;'' the active component is propagated by a space-specific effective Hamiltonian different from the rest. The vertex corrections are introduced by a rescaled time-dependent nonlocal exchange interaction. The direct $\Gamma$ correction to the self-energy is further updated by adjusting the rescaling factor in a self-consistent post-processing cycle. Our embedding method is tested mainly on donor-acceptor charge-transfer systems. The embedded vertex effects consistently and significantly correct the quasiparticle energies of the gap-edge states. The fundamental gap is generally improved by 1-3 eV upon the one-shot $GW$ approximation. Furthermore, we provide an outlook for applications of (embedded) vertex corrections in calculations of extended solids.
\end{abstract}

\maketitle

\section{Introduction}
Predicting optoelectronic properties of functional materials hinges upon the availability of accurate and efficient first-principles methods.\cite{Goedecker1999,Goedecker2003,Skylaris2005,VandeVondele2005,Zhou2006,Neese2009,Saad2010,Baer2013,Neuhauser2014} Particularly, excitation energies obtained from electron structure calculations are directly related to the charge-transfer processes and optical transitions in practical materials. This requires the ability to capture excited-state properties, especially the nonlocal and dynamical electron-electron interactions. Many-body perturbation theory (MBPT) within the Green's function formalism\cite{fetter2003,Martin2016} provide a powerful tool for solving electron correlation problems. The $GW$ approximation\cite{Hedin1965,Aryasetiawan1998,Onida2002,Schindlmayr2006,Golze2019} assumes that the many-body interactions (nonlocal and dynamical electron-electron exchange and correlation) are represented by the exchange-correlation self-energy: $\Sigma_{\rm{XC}} = {\rm i}GW$, which is a convolution of the single-particle excitation propagator, i.e., Green's function ($G$), and the screened Coulomb interaction ($W$). In practice, the correlation stems from nonlocal and time-dependent charge density-density interactions. The $GW$ method provides direct access to quasiparticle (QP) energies and has been widely applied to molecular and condensed systems.\cite{Strinati1980,Strinati1982,Hybertsen1985,Hybertsen1986,Godby1988,Aulbur2000,Rostgaard2010,Huser2013,vanSetten2013,Korbel2014,Chen2014,Govoni2015,Gallandi2015,Caruso2016,Kang2016,Rangel2016,Gallandi2016,Hung2017,Bois2017,Govoni2018,Ergonenc2018} The emerging stochastic $GW$ approach\cite{Neuhauser2014_prl,Vlcek2017,Vlcek2018_prb} further enables its application to systems with up to thousands of electrons.\cite{Vlcek2018_prm,Brooks2020,Weng2020,Romanova2020,Weng2021,Romanova2022} In these approaches, the QP energy is obtained by applying the $\Sigma_{\rm{XC}}$ as a perturbative correction to the mean-field eigenvalue, typically computed using Kohn-Sham density functional theory (KS-DFT).\cite{Hohenberg1964,Kohn1965}

Although significant improvement upon the mean-field results has been achieved, the $GW$ approximation suffers from large errors in predicting QP energies of unoccupied states\cite{Rangel2016} and fails to capture the satellite peaks in the photoemission spectra.\cite{Cederbaum1977,Cederbaum1980} These failures can be attributed (at least in part) to the self-polarization error.\cite{Nelson2007,Aryasetiawan2012} They are remedied by the inclusion of the vertex correction, which is completely neglected by the $GW$ approximation. Only recently, vertex-corrected methods have started to emerge in practical calculations.\cite{Del1994,Tiago2006,Morris2007,Shishkin2007,Gruneis2009,Chang2012,Gruneis2014,Ren2015,Chen2015,Hung2016,Kuwahara2016,Knight2016,Schmidt2017,Maggio2017,Hellgren2018_prb,Vlcek2019,Lewis2019,Ma2019,Pavlyukh2020,Zaera2021,Wang2021,Wang2022} The vertex corrections are applied to two places of the $\Sigma_{\rm{XC}}$: one enters the irreducible polarizability in the $W$ term; the other is a term explicitly included in the expression of $\Sigma_{\rm{XC}}$. Within the random phase approximation (RPA), the $W$ term is computed using the independent particle picture, where $P = -{\rm i}GG$ ($P$ denotes the polarizability). The $P$ computed by RPA neglects electron-hole ladder diagrams and higher-order interactions for the $W$ term. The vertex correction, $\Gamma$, to the polarizability amounts to $P = -{\rm i}GG\Gamma$, which is in principle an exact formulation for $P$. This leads to the so-called $GW^{tc}$ approximation.\cite{Del1994,Bruneval2005,Romaniello2009} The vertex function $\Gamma$ entering the $\Sigma_{\rm{XC}}$ leads to $\Sigma_{\rm{XC}} = {\rm i}GW^{tc}\Gamma$. Due to the intractability of the full vertex function, low-order\cite{Freeman1977,Del1994,deGroot1996,Gruneis2009,Ren2015,Knight2016,Kuwahara2016,Lewis2019,Pavlyukh2020} and local vertex\cite{Romaniello2009,Tiago2006,Morris2007,Romaniello2012,Aryal2012,Hung2016,Schmidt2017,Hellgren2018_prb,Hellgren2018} approximations are often employed. The vertex correction to the $W$ term is shown to minimally improve the $GW$ results for molecular ionization potentials (IP).\cite{Lewis2019} In contrast, electron affinity (EA) predictions are enhanced in several cases.\cite{Knight2016,Schmidt2017,Vlcek2019} Fundamentally, the inclusion of vertex in the self-energy (i.e., the $GW^{tc}\Gamma$ approach) successfully captures multi-quasiparticle interactions and satellite peaks in molecular photoemission spectra of all valence electrons,\cite{Zaera2021,Zaera2022} agreeing excellently with the adaptive sampling (nearly full) configuration interaction approach.\cite{Tubman2016,Tubman2018,Zaera2019,Tubman2020}

Despite the success of $GW^{tc}\Gamma$ in molecules, extending it to condensed systems, e.g., solids or low-dimensional materials, is still challenging. The polarizability evaluated with electron-hole ladders can be paralleled with employing a Hamiltonian with a nonlocal exchange interaction in the time-dependent density functional theory (TDDFT) calculation. The spatial nonlocality of the electron-electron interactions is (to the lowest order) analogous to the Hartree-Fock (HF) approximation or the generalized Kohn-Sham (GKS) DFT. Furthermore, this nonlocal exchange should be screened corresponding to the nonlocal correlation, which is especially critical for semiconducting systems. A practical approach is to use a statically screened exchange interaction in parallel with the optimally-tuned GKS functionals.\cite{Stein2009_jcp,Stein2010,Kronik2012,Abramson2012,Abramson2015,Manna2018,Kronik2018,Bhandari2018,Wing2019,Prokopiou2022} This approach has been successfully used as the $GW$ starting points for molecules,\cite{Gallandi2015,Gallandi2016,Rangel2016,Knight2016,Brawand2016,Bois2017,Rangel2017} and the optimized GKS Hamiltonian often yields excellent optical spectra.\cite{Stein2009,Abramson2015,Manna2018,Wing2019} However, the optimal tuning of the range-separation parameter is non-trivial for extended (periodic) systems.\cite{Vlcek2015,Wing2021,Gant2022,Ohad2022} In addition, the optimally tuning process based on the Koopman's\cite{Koopmans1934} or Janak's\cite{Janak1978} theorem involves only frontier states. The resulting parametrization is not necessarily ``optimal'' for other electronic states (e.g., in the bottom valence region). For the nonlocal vertex in extended systems, several attempts have been explored to compute the exchange-correlation kernel, including the use of ad hoc\cite{Shishkin2007,Gruneis2014} (e.g., HSE) or dielectric-dependent\cite{Ma2019} hybrid functionals and the approximate bootstrap iteration.\cite{Chen2015}

To tackle the computational cost of including the nonlocal vertex in large extended systems, we propose an efficient embedding method that includes vertex corrections in the $G_0W_0$ correlation self-energy (here computed using stochastic sampling). We use an active space projector\cite{Romanova2020,Weng2021} to separate electronic states into two components: the active part and its orthogonal complement denoted as the ``rest.'' The active space projector is composed of either canonical KS states or localized orbitals that are energetically favored in optical transitions. In this scheme, the electron-hole interactions are selectively ``turned on'' for the active space. A \textit{space-specific} effective Hamiltonian is constructed with a stochastically sampled and \textit{rescaled} time-dependent nonlocal exchange interaction. The active space component is specifically propagated using this effective Hamiltonian, while the evolution of the rest is treated by a mean-field one. The recombination of these two components produces the embedded states that enter the correlation self-energy. Furthermore, we propose a simple self-consistent post-processing cycle for rescaling the vertex contribution.

The proposed embedding method is mainly tested on $\pi$-conjugated donor-acceptor systems with significant electron-hole interactions. The investigated systems include isolated small molecules and donor-acceptor dimers, one-dimensional (1D) charge-transfer copolymer, and two-dimensional (2D) donor-acceptor double layers. Different sizes and representations of the active space are also explored. The embedded vertex corrections show consistent and non-trivial effects on the computed QP energies. We find that the vertex correction to the polarizability is critical to avoid overbound unoccupied states. The fundamental gaps are 1-3 eV greater than the $G_0W_0$ ones and hence improving upon the $G_0W_0$ results. More importantly and in contrast with previous findings, the vertex corrections significantly affect the QP energies of the occupied gap-edge states.

The following content contains three major parts: Section \ref{sec:tandm} focuses on the theory and methodology; Section \ref{sec:resdisc} presents the computed charge excitation energies and the fundamental gaps of various systems; Section \ref{sec:candp} summarizes the findings and comments on possible further improvement.

\section{Theory and Methodology}\label{sec:tandm}
In this section, the concept of QP self-energy is revisited and various forms of vertex-corrected self-energy are introduced. Second, we demonstrate the definition and construction of an active space. Finally, the vertex-embedding scheme is presented. 
\subsection{Quasiparticle Self-Energy}\label{sec:QPandSE}
The QP Hamiltonian is written as
\begin{equation}
   \hat{H}^{QP} = \hat{T} + \hat{V}_{\rm{ext}} + \hat{\Sigma}_{\rm{H}} + \hat{\Sigma}_{\rm{XC}}(\omega),
\label{eqn:QPH}
\end{equation}
where $\hat{T}$, $\hat{V}_{\rm{ext}}$, and $\hat{\Sigma}_{\rm{H}}$ represent the kinetic energy operator, external field potential, and classical Coulomb repulsion, respectively. All nonlocal and dynamical particle interactions are included in the frequency-dependent exchange-correlation self-energy $\hat{\Sigma}_{\rm{XC}}$. Next, we separate the exchange and correlation interactions into two individual terms and focus on the correlation part. The perturbatively-corrected QP energy reads
\begin{equation}
   \varepsilon_j^{QP} = \braket{\phi_j |\varepsilon_j^{0} - \hat{v}_{\rm{xc}} + \hat{\Sigma}_{\rm{X}} + \hat{\Sigma}_{\rm{C}}(\omega=\varepsilon_j^{QP}) | \phi_j}.
\label{eqn:QPE}
\end{equation}
Here, $\phi_j$ is an eigenstate of the KS-DFT Hamiltonian with the corresponding eigenvalue $\varepsilon_j^{0}$ and exchange-correlation potential $\hat{v}_{\rm{xc}}$. As a starting point, our DFT calculations employ the PBE functional.\cite{Perdew1996} Other starting points are also applicable but out of the scope of this work since we focus on demonstrating the embedding scheme. The nonlocal exchange $\hat{\Sigma}_{\rm{X}}$ is equivalent to the Fock operator in the Hartree-Fock approximation.\cite{Martin2016} The frequency-dependent correlation self-energy $\hat{\Sigma}_{\rm{C}}(\omega)$ is obtained from the Fourier-transformed time-dependent $\hat{\Sigma}_{\rm{C}}(t)$, which is approximated and computed by the stochastic methods.\cite{Neuhauser2014_prl,Vlcek2018_prb,Vlcek2019} Only diagonal terms of the self-energy are considered.

In the $GW^{tc}\Gamma$ formalism, the temporally and spatially nonlocal correlation self-energy ($\Sigma_{\rm{C}}$) is derived from the derivative of the total self-energy ($\Sigma_{\rm{T}}$) with respect to the Green's function ($G$).\cite{Vlcek2019,Zaera2021,Zaera2022} The total self-energy reads
\begin{equation}
  \Sigma_{\rm{T}} = \Sigma_{\rm{H}} + \Sigma_{\rm{X}} + \Sigma_{\rm{C}}.
\label{eqn:totalse}
\end{equation}
Since the right-hand side (RHS) of Eq.~\eqref{eqn:totalse} also contains the $\Sigma_{\rm{C}}$ term, the correlation self-energy thus needs to be solved self-consistently. Here, we expand the simplified approach presented in previous work\cite{Maggio2017,Vlcek2019}, where the variation of $\Sigma_{\rm{C}}$ is neglected. We derive the vertex ($\Gamma$) from a statically rescaled nonlocal exchange interaction. Hence, the exchange and correlation terms in Eq.~\eqref{eqn:totalse} are combined as
\begin{equation}
  \Sigma_{\rm{X}} + \Sigma_{\rm{C}} \approx \beta \Sigma_{\rm{X}},
\label{eqn:rescalehf}
\end{equation}
where $\beta$ is simply a prefactor multiplied by the nonlocal exchange interaction. The \textit{rescaling} factor $\beta$, derived in section~\ref{sec:screen}, is similar to the state-dependent screened exchange constant used in previous work.\cite{Brawand2016}

Based on the approximation above (Eq.~\eqref{eqn:rescalehf}), the vertex is introduced in a greatly simplified way that preserves the spatial nonlocality of $\Gamma$ (full derivation is provided in the supplemental material). In the time-dependent formalism, it is convenient to decompose the correlation self-energy into two terms:
\begin{equation} 
\Sigma_{\rm{C}}^{G_0W_0^{tc}\Gamma_{\beta}}(t) = \Sigma_{\rm{C}}^{G_0W_0^{tc}}[\delta n(t), t]  + \Sigma_{\rm{C}}^{\Gamma_{\rm{\beta}}}[\delta \rho(t), t].
\label{eqn:corln}
\end{equation}
The first term on the RHS of Eq.~\eqref{eqn:corln} comes from the derivative of the Hartree term ($\Sigma_{\rm{H}}$) with respect to the Green's function ($G$), and it is a functional of the induced density $\delta n$. This is nothing else but the $G_0W_0^{tc}$ self-energy representing the correlation stemming from density-density interactions. Here, the vertex correction is introduced to the $W$ by applying a Hamiltonian with a time-dependent nonlocal exchange interaction in the propagation process (see section~\ref{sec:evc}). The second contribution, $\Sigma^{\Gamma_{\beta}}_{\rm{C}}$, comes from the derivative of the rescaled nonlocal exchange ($\beta \Sigma_{\rm{X}}$) with respect to the Green's function ($G$). In practice, it encodes the correlation due to density matrix fluctuations is thus a functional of the induced density matrix ($\delta \rho$). Since $\beta$ is simply a prefactor, we extract it from the derivative and have the following
\begin{equation}
 \Sigma_{\rm{C}}^{\Gamma_{\beta}}[\delta \rho(t), t] = \beta \Sigma_{\rm{C}}^{\Gamma_{\rm{X}}}[\delta \rho(t), t]
\label{eqn:rsgama},
\end{equation}
where $\Sigma_{\rm{C}}^{\Gamma_{\rm{X}}}$ is derived from a bare $\Sigma_{\rm{X}}$ in this work. In the following sections, we use $\Gamma_{W}$ to denote the vertex correction to the $W$ term (i.e., the polarizability) and the direct vertex correction to the self-energy is denoted as either $\Gamma_{\rm{X}}$ or $\Gamma_{\beta}$ depending on the precise definition detailed below.

\subsection{Electronic Active Space}\label{sec:actsp}
In the real-time $G_0W_0$ formalism, the induced density is computed by the time-dependent Hartree (TDH) approximation, which corresponds to RPA. The TDH Hamiltonian does not couple the electron-hole pairs. Upon excitation, electron and hole states at low excitation energies are preferably populated and are expected to contribute predominantly to the polarizability. The gap-edge states, i.e., the highest occupied molecular orbital (HOMO) and the lowest occupied molecular orbital (LUMO), are mostly responsible for electron-hole pair formation captured by the ladders in vertex-corrected polarizability (detailed in the next section). For $\pi$-conjugated molecules, the $\pi$-$\pi^*$ transition is considered the governing element in the optical absorption spectrum. $\pi$ and $\pi^*$ bonds thus represent another type of highly-populated electron-hole states. 

Here, we first identify the low-energy states (e.g., HOMO/LUMO or $\pi$/$\pi^*$). An electronic active space is then formed by these states based on their population (occupation) during excitation, and we denote its orthogonal complement as the ``rest.'' The active space is then treated differently from the rest (see the next section).

Mathematically, the active space is defined by
\begin{equation}
   \hat{P}^A = \sum_i^{N_{\rm act}} \ket{\psi_{i}} \bra{\psi_{i}},
\label{eqn:activep}
\end{equation}
where $\hat{P}^{A}$ is a projector onto the active space, and $N_{\rm act}$ represents the number of projector states $\psi$.

This work explores two different representations of $\psi$: canonical KS eigenstates and localized orbitals. The first representation straightforwardly chooses canonical eigenstates, e.g., the HOMO and the LUMO, based on their mean-field eigenvalues. In contrast, the localized basis allows the selection of a specific type of chemical bonds, especially the $\pi$ and $\pi^*$ bonds mentioned above. We employ Pipek-Mezey (PM) localized orbitals\cite{Pipek1989} because they easily separate the $\sigma$ and $\pi$ characters.\cite{Pipek1989,Jonsson2017} Analogously, for periodic systems, the top valence band (TVB) and the bottom conduction band (BCB) provide the gap-edge states, which can be represented by either Bloch states (canonical eigenstates) or combinations of Wannier functions. The regionally localized orbitals for molecular systems are obtained using the sequential PM localization on the chosen molecules.\cite{Weng2022}. In the next section, we demonstrate the embedding of vertex corrections using the defined active space.

\subsection{Embedded Vertex Corrections}\label{sec:evc}
This section demonstrates the embedded vertex corrections in the correlation self-energy (Eq.~\eqref{eqn:corln}). In the $GW$ approximation, the polarizability computed by RPA includes only the electron-hole ``bubble'' diagrams but neglects higher-order interactions. In the following, we go beyond the $GW$ approximation and selectively turn on electron-hole interactions (vertex in the polarizability) for the active space by applying a space-specific effective Hamiltonian, which governs the real-time dynamics mimicking the time-dependent solution to the Bethe-Salpeter equation\cite{Rabani2015}
\begin{equation}
  \hat{H}^{\rm eff}(t) = \hat{T} + \hat{V}_{\rm{ext}} + \hat{\Sigma}_{\rm{H}}[n(t)] + \alpha \hat{\Sigma}_{\rm{X}}[\rho(t)].
\label{eqn:eQPH}
\end{equation}
The rest of the electronic Hilbert space is instead treated by 
\begin{equation}
  \hat{H}^{0}(t) = \hat{T} + \hat{V}_{\rm{ext}} +  \hat{v}_{\rm{xc}} + \hat{\Sigma}_{\rm{H}}[n(t)].
\label{eqn:mfH}
\end{equation}
Compared to the $\hat{H}^0$, the $\hat{H}^{\rm eff}$ contains a rescaled time-dependent nonlocal exchange interaction, constituting the lowest-order vertex correction to the polarizability. The coefficient $\alpha$ is another rescaling factor analogous to the screening factor. Similar to the $\beta$ defined above, it is derived from the expectation values of the underlying self-energy (see section~\ref{sec:screen}). The $\alpha$ factor represents the averaged rescaling behavior for the entire active space, while $\beta$ is used for a specific state.

In practice, the Hamiltonian separation is employed in the real-time evolution of electronic states yielding the time-dependent induced density $\delta n(t)$ and density matrix $\delta \rho(t)$ in Eq.~\eqref{eqn:corln}. To make this approach efficient even for large systems, we employ the stochastic approach that samples these two quantities using a set of random vectors instead of the deterministic single-particle states. Technical details of the stochastic formalism are provided in the supplemental material. Here, we only emphasize the quantities that are directly related to the embedding.

\begin{figure}
    \centering
    \includegraphics[width=0.45\textwidth]{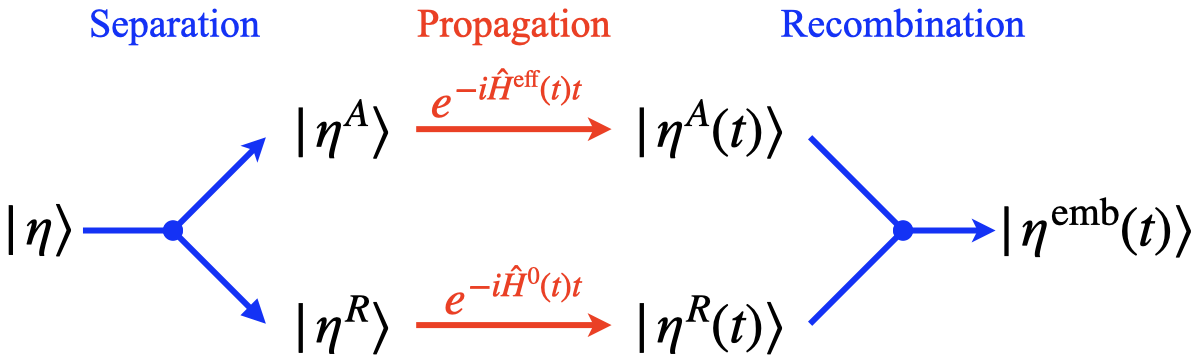}
    \caption{Schematic representation of the separation-propagation-recombination technique for the embedded time-evolved random vectors. The separation and recombination steps are colored in blue and the propagation step is colored in red.}
    \label{fig:spr_scheme}
\end{figure}

The time-dependent density $n(t)$ and density matrix $\rho(t)$ are sampled by a set of random states $\eta$ (defined in Eqs.~\eqref{eqn:zetabar} and~\eqref{eqn:eta})
\begin{equation}
    \label{eqn:embdens}
    n(\textbf{r},t) = \{ \eta^*(\textbf{r},t) \eta(\textbf{r},t) \},
\end{equation}
and 
\begin{equation}
    \label{eqn:embdensmat}
    \rho(\textbf{r}_1, \textbf{r}_2, t) = \{ \eta^*(\textbf{r}_2,t) \eta(\textbf{r}_1,t) \}.
\end{equation}
Here, the brackets $\{\dots\}$ denote an average over the whole set of random functions. The time-evolution of $\eta$ gives $n(t)$ and $\rho (t)$, which produce the $\delta n(t)$ and $\delta \rho(t)$ for Eq.~\eqref{eqn:corln}. The core procedure of vertex-embedding is to prepare $\eta^{\rm emb}$ by the following \textit{separation-propagation-recombination} (SPR) treatment (Fig.~\ref{fig:spr_scheme}).

At $t=0$, the random vector $\eta$ is projected onto the occupied subspace. The electron removal/addition rotates the $\eta$ to the full Hilbert space, i.e., it also represents electrons excited into the unoccupied subspace. The component in the active space, $\eta^A$, of the perturbed $\eta$ is obtained by projection following Eq.~\eqref{eqn:activep}
\begin{equation}
    \label{eqn:acteta}
    \ket{\eta^A} = \hat{P}^{A} \ket{\eta} = \sum_i^{N_{\rm act}} \braket{\psi_i|\eta} \ket{\psi_{i}}.
\end{equation}
The rest component $\eta^R$ is then given by
\begin{equation}
    \label{eqn:resteta}
    \ket{\eta^R} = \ket{\eta} - \ket{\eta^A}.
\end{equation}
The two separated components are propagated using two different operators
\begin{equation}
  \ket{\eta^A(t)} = \hat{U}^A(t) \ket{\eta^A},
\label{eqn:astate}
\end{equation}
and
\begin{equation}
  \ket{\eta^R(t)} = \hat{U}^R(t) \ket{\eta^R}.
\label{eqn:rstate}
\end{equation}
The two time-evolution operators above correspond to the two forms of Hamiltonian in Eqs.~\eqref{eqn:eQPH} and~\eqref{eqn:mfH}, which (adiabatically) depend on the density (and density matrix) at time $t$
\begin{equation}
  \hat{U}^A(t) = e^{-{\rm i}\hat{H}^{\rm eff}(t)t},
\label{eqn:activeu}
\end{equation}
and
\begin{equation}
  \hat{U}^R(t) = e^{-{\rm i}\hat{H}^0(t)t}.
\label{eqn:restu}
\end{equation}
Finally, the embedded time-evolved random vector $\eta^{\rm emb}(t)$ is given by recombining the two components
\begin{equation}
  \ket{\eta^{\rm emb}(t)} = \ket{\eta^A(t)} + \ket{\eta^R(t)}.
\label{eqn:addtn}
\end{equation}

The SPR scheme above is demonstrated in a two-component case, and the extension to three or more components is trivial. The three-component case, in which the active space is further divided into the occupied and unoccupied parts, is applied in this work as shown below. In practice, the propagation in time is discretized. In each time step, the separation treatment is repeated to ensure $\eta^{A}(t)$ and $\eta^{R}(t)$ are orthogonal before they are propagated. Please refer to Eqs.~\eqref{eqn:eactcom2}-\eqref{eqn:etotaleta} in the supplemental material for more details.

The embedded vector $\eta^{\rm emb}(t)$ generates the embedded density $n^{\rm emb}(t)$ via Eq.~\eqref{eqn:embdens} and density matrix $\rho^{\rm emb}(t)$ via Eq.~\eqref{eqn:embdensmat}. Finally, the $n^{\rm emb}(t)$ and $\rho^{\rm emb}(t)$ lead to the embedded self-energy (Eq.~\eqref{eqn:corln}). The time-evolution behavior of $\eta$ is modified, but the total number of electronic states sampled is conserved. And therefore, there is no double-counting error in this embedding scheme.

The SPR procedure also applies to the random vectors that sample the Green's function $G_0$. In this case, the time evolution is simplified as $G_0$ is the non-interacting propagator depending on a static form of the Hamiltonian. More details are provided in the ``Embedded Vertex Correction'' section of the supplemental material.

\subsection{Rescaling Factors}\label{sec:screen}
In the following, we comment on the working definition of the two rescaling factors: $\beta$ in Eq.~\eqref{eqn:rescalehf} and $\alpha$ in Eq.~\eqref{eqn:eQPH}. In practice, they are derived simply as a fraction of the nonlocal exchange interaction that mimics the expectation value of the full exchange-correlation self-energy. In particular, we first derive them from the $G_0W_0$ approximation though; as we show below, they can be generalized to a simple self-consistent post-processing approach.

For a QP state $\phi_j$, the initial rescaling factor $\beta_j^{0}$ is given by
\begin{equation}
    \beta_j^{0} = \frac{\Sigma_{\rm{X},j}+ \Sigma^0_{\rm{C}}(\omega = \varepsilon^{QP}_j)}{\Sigma_{\rm{X},j}},
\label{eqn:betaj}
\end{equation}
where $\Sigma_{\rm{X},j} = \braket{\phi_j|\hat{\Sigma}_{\rm X}|\phi_j}$, and the correlation self-energy $\Sigma^0_{\rm{C}}$ is an expectation value computed at the $G_0W_0$ level. Note that due to the definition of $\Sigma_{\rm{X}}$ in Eq.~\eqref{eqn:QPE}, the value of $\Sigma_{\rm{X},j}$ is always negative throughout this work. The computed $\Sigma_{\rm{C}}$, as shown in the Results and Discussion section, can have either the same or an opposite sign with/against $\Sigma_{\rm{X},j}$. In the same-sign case, the derived $\beta_j$ from Eq.~\eqref{eqn:betaj} is greater than 1, and we define this effect as correlation stabilization. The opposite-sign case gives $\beta_j<1$, which is instead defined as correlation destabilization.

For the rescaling factor $\alpha$ (Eq.~\eqref{eqn:eQPH}) that represents the entire active space, we first compute the $\beta_i^0$ of each state $\psi_i$ in Eq.~\eqref{eqn:activep} via Eq.~\eqref{eqn:betaj}. The $\alpha$ for the entire active space takes the average value of $\beta_i^0$s
\begin{equation}
    \alpha = \frac{1}{N_{\rm act}} \sum^{N_{\rm act}}_{i=1} \beta_i^{0}.
\label{eqn:alpha}
\end{equation}
As shown in the Results and Discussion section, the states $\psi_i$ within one active space have very similar character and consistent rescaling behavior. It is thus sufficient to use a single averaged $\alpha$ to describe the entire active space. From the correlation self-energy perspective, the parameters $\alpha$ and $\beta$ play two different roles: the $\alpha$ representing an active space enters the effective Hamiltonian in Eq.~\eqref{eqn:eQPH} that governs the time evolution and determines the electron-hole coupling strength; the $\beta_j$ of a state $\phi_j$ in Eq.~\eqref{eqn:QPE} rescales $\Sigma_{\rm X}$ and thus the $\Sigma^{\Gamma_{\rm{X}}}_{\rm{C}}$ term in Eq.~\eqref{eqn:rsgama}.

\begin{figure}
    \centering
    \includegraphics[width=0.2\textwidth]{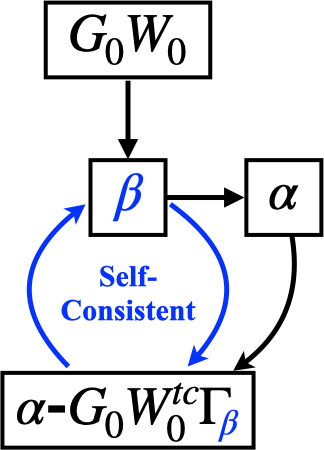}
    \caption{Schematic representation of the self-consistent $\alpha$-$G_0W_0^{tc}\Gamma_{\beta}$ correlation self-energy. The self-consistent cycle is colored in blue.}
    \label{fig:sc_scheme}
\end{figure}

With the $\alpha$ and $\beta_j^{0}$ computed by $G_0W_0$, we solve the vertex-corrected self-energy in Eq.~\eqref{eqn:corln}. $\beta_j$ acts as a factor only in front of the $\Sigma^{\Gamma_{\rm{X}}}_{\rm{C}}$ (see Eq.~\eqref{eqn:rsgama}). We thus compute the $\Sigma^{\Gamma_{\rm{X}}}_{\rm{C}}(t)$ only once and use it to update $\beta_j$ through the self-consistent cycle in Fig.~\ref{fig:sc_scheme}
\begin{equation}
  \beta_j^{n} = \frac{\Sigma_{\rm{X},j} + \Sigma^{n-1}_{\rm{C}}[\beta_j^{n-1}] (\omega = \varepsilon^{QP}_j)}{\Sigma_{\rm{X},j}},
\label{eqn:scbetaj}
\end{equation}
where $\Sigma^{n-1}_{\rm{C}}$ is the vertex-corrected correlation self-energy using the prefactor $\beta_j^{n-1}$ in Eq.~\eqref{eqn:corln}. Note that this self-consistency comes at a negligible computational cost as $\beta_j$ only rescales one (additive) component of the self-energy. In principle, the factor $\alpha$ can also be updated self-consistently, but this instead requires re-evaluating the time evolution. We avoid this step for simplicity here.

In the remainder of this work, $\alpha$-$G_0W_0^{tc}$ denotes the correlation self-energy with only $\Gamma_{W}$ correction; $\alpha$-$G_0W_0^{tc}\Gamma_{\beta^0}$ and $\alpha$-$G_0W_0^{tc}\Gamma_{\beta^{sc}}$ represent the correlation self-energy with both $\Gamma_{W}$ and $\Gamma_{\beta}$ corrections, and the latter includes self-consistency in $\beta_j$.

\section{Results and Discussion}\label{sec:resdisc}
In the following sections, we demonstrate and test the performance of the proposed methodology on selected practically important materials. Specifically, we focus on calculating QP energies of the gap-edge states for molecules and band-edge states for periodic systems. We choose to study isolated and low-dimensional (1D and 2D) periodic systems so that for each case we can determine the absolute QP energy with respect to vacuum (in contrast, for 3D periodic systems only QP energy differences are meaningful). Investigated systems include tetracyanoethylene (TCNE) and tetracyanoquinodimethane (TCNQ) acceptors, 1,4-benzenediamine-TCNE (B-T) dimer, tetracene-C$_{60}$ (Tc-C) dimer, poly-fluorene-benzothiadiazole (p-FBT) polymer, and tetracene-C$_{60}$ (Tc-C) double layers. The structures of the isolated molecules, molecular dimers and, polymer are prepared or taken from Refs.~\citenum{Zhang2014,Weng2020,Mei2020}. The double-layer system is constructed by using the dimer geometry with the periodic boundary conditions of the C$_{60}$ crystal. 

For isolated systems, the active space is constructed by either canonical frontier orbitals, i.e., the HOMO and LUMO, or $\pi$/$\pi^*$ bonds. For periodic systems, the active space is represented by either Bloch states or Wannier functions. The stochastic $G_0W_0$ and $\alpha$-$G_0W_0^{tc}$ results are reported with $<$0.05 eV statistical errors. The statistical erros for the $\alpha$-$G_0W_0^{tc}\Gamma_{\beta}$ results are slightly larger ($<$0.07 eV) due to additional fluctuations in sampling the induced density matrix. All the numeric results in this work are supported by the graphical solutions on the self-energy curves (see Figs.~\ref{fig:tcne_eqp}-\ref{fig:tc_dl_eqp} in the supplemental material).

\subsection{Single Acceptor Molecules}
\begin{figure}
    \centering
    \includegraphics[width=0.4\textwidth]{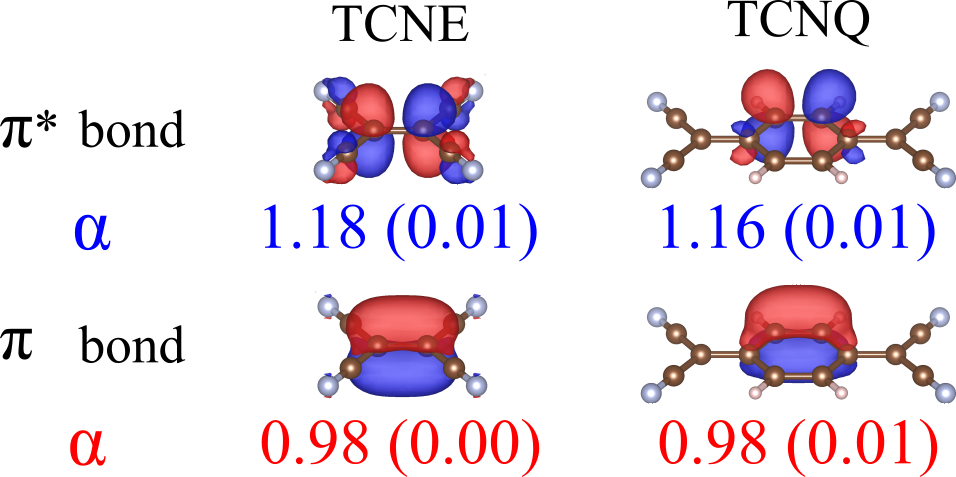}
    \caption{Representative $\pi$ and $\pi^*$ bonds of the TCNE and TCNQ molecules. The rescaling factor is averaged over 9 and 12 states for the TCNE and TCNQ, respectively, with the standard deviations provided in parenthesis.}
    \label{fig:acceptor}
\end{figure}

TCNE and TCNQ are ideal molecules to benchmark EA predictions due to their strongly bound LUMO.\cite{Ren2015,Knight2016,Zakrzewski1996} To test the performance of our method on these two acceptors, we use the $\pi$+$\pi^*$ active space. For TCNE, nine $\pi$ bonds and nine $\pi^*$ bonds are identified from the PM localized orbitals (Fig.~\ref{fig:acceptor_basis}). The computed rescaling factors $\alpha$ and the representative orbitals for the active occupied and unoccupied spaces are listed in Fig.~\ref{fig:acceptor}. The ratio $\Sigma_{\rm{XC}} / \Sigma_{\rm{X}}$ ($\alpha$) is smaller than 1 for the $\pi$ bonds but greater than 1 for the $\pi^*$ bonds, i.e., the value of $\Sigma_{\rm{C}}$ is positive for occupied states but negative for unoccupied states. This illustrates the role of the correlation self-energy component, which attenuates the nonlocal exchange and destabilizes occupied states but leads to the energetic stabilization of unoccupied states. The latter corresponds to the apparent ``strengthening'' of the nonlocal exchange interaction, and the $\alpha$ is counterintuitively greater than 1. These rescaling factors $\alpha$ are used in the time evolution (Eq.~\eqref{eqn:eQPH}) for the active space component. Note that this approach attempts to include the correlation via rescaling the exchange interaction $\Sigma_{\rm{X}}$ so that the QP energies are reproduced. Such an approach is undoubtedly a crude simplification, yet it enables an efficient proof of principle calculations described here; we comment on the shortcomings of this implementation in the following.

The computed HOMO/LUMO energies and fundamental gaps are summarized in Table~\ref{tab:acceptor}. Our $G_0W_0$ QP energies agree perfectly with previously reported results,\cite{Ren2015,Knight2016} and the corresponding HOMO/LUMO gap increases to 7.38 eV. When the $\Gamma_{W}$ is applied, the gap is further corrected by $\alpha$-$G_0W_0^{tc}$: the HOMO energy is shifted down by $\sim$0.6 eV, and the LUMO energy is shifted up by $\sim$1 eV. These corrections to IP and EA are more significant than the comparable $G_0W_0$+SOSEX\cite{Ren2015,Knight2016} approach (see Tables~\ref{tab:ip_compare} and~\ref{tab:ea_compare}). And the corrected results agree better with experiments,\cite{Houk1976,Lyons1976} especially with the electron affinity. We note that the 2.30-eV result in Table~\ref{tab:acceptor} cited from Ref.~\citenum{Lyons1976} represents the \textit{vertical} EA,\cite{Zakrzewski1996} while the \textit{adiabatic} EA from experiments is $\sim$3.16 eV.\cite{Chowdhury1986,Zakrzewski1996,Khuseynov2012,Ren2015} Our results are compared with the former since the QP energies correspond to vertical charge excitations. The $G_0W_0$ result highly overestimates the EA by $\sim$1.5 eV, and the inclusion of $\Gamma_{W}$ improves it by $\sim$1 eV. In other words, the vertex correction to the polarizability is providing a significant improvement. This contrasts some previous observations,\cite{Lewis2019,Vlcek2019} indicating that the effect of $\Gamma_{W}$ is not universal.  

When $\Gamma_{\beta}$ is added, the HOMO level is stabilized by $\sim$1 eV, and the $\beta$-self-consistency contributes another $\sim$0.1 eV. Further, the inclusion of $\Gamma_{\beta}$ shifts the LUMO level up to a higher energy. However, the quantitative effect appears too strong when using the rescaled nonlocal exchange approach: when $\Gamma_{\beta}$ is applied with $\beta=1$, i.e., $\Gamma_{\rm{X}}$, the LUMO energy is strongly destabilized to $-$0.45 eV. The result worsens with $\beta>1$, as suggested by the oversimplified approach based on the ratio of the self-energy expectation values. The nearly unbound value is incorrect and is against the acceptor nature of TCNE and the experiment. The derived $\Gamma_{\beta}$ from a simple rescaled exchange interaction fails for the LUMO.

\begin{figure}
    \centering
    \includegraphics[width=0.4\textwidth]{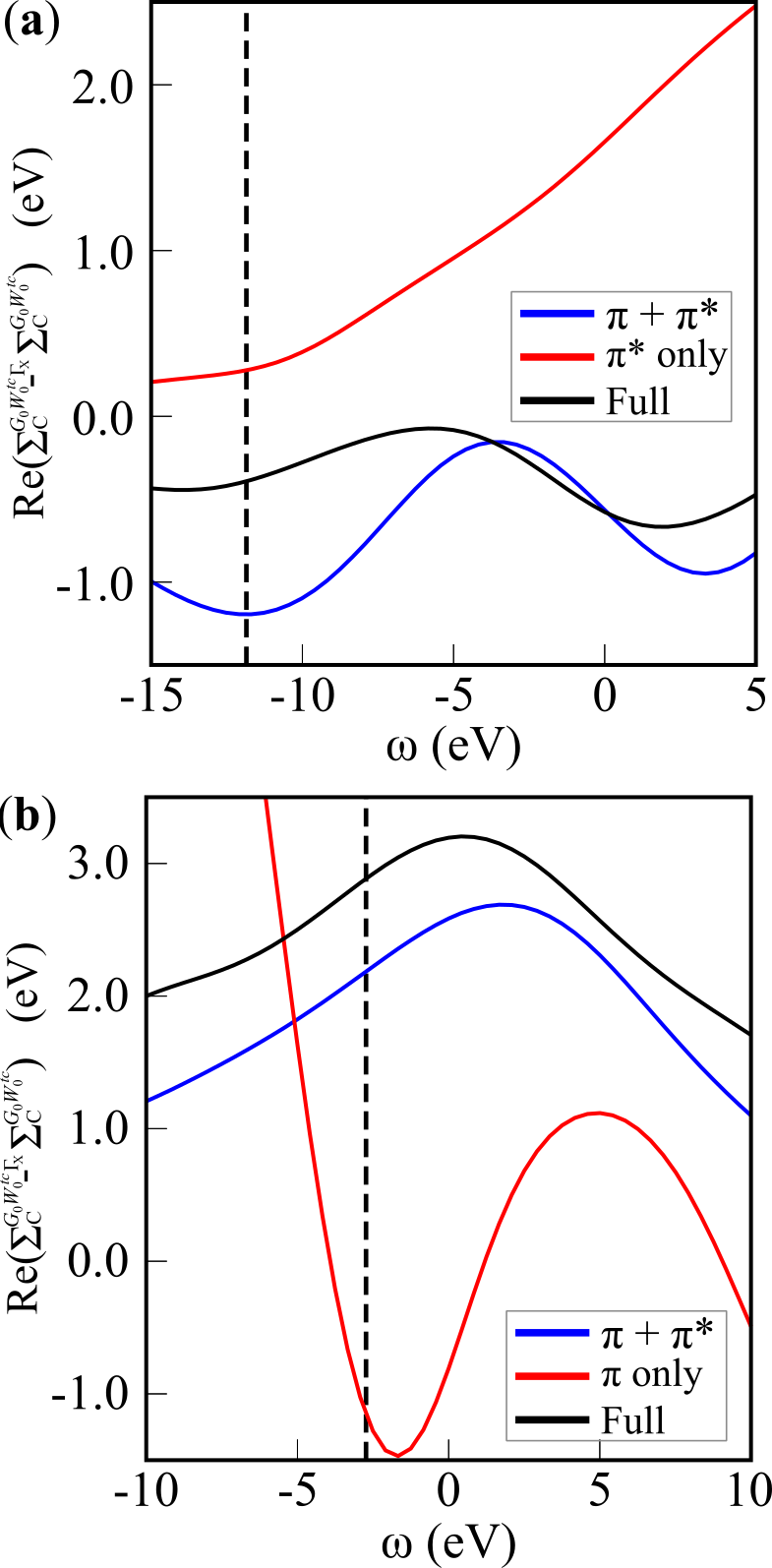}
    \caption{Real parts of the $\Sigma^{\Gamma_{\rm{X}}}_{\rm{C}}$ calculated for the (a) HOMO and the (b) LUMO using different active spaces. The black curve represents the full Hilbert space, the red curve denotes the $\pi$ or $\pi^*$ only space, and the blue curve is the $\pi+\pi^*$ active space. The dashed line indicates the QP energy from the $\alpha$-$G_0W_0^{tc}$ results.}
    \label{fig:tcne_se}
\end{figure}

To further understand the $\Gamma_{\beta}$ contribution, we neglect the rescaling effect by setting $\beta=1$ and examine the real part of the frequency-dependent $\Sigma_{\rm{C}}^{\Gamma_{\rm{X}}}$. By varying the components of the active space, we obtain several $\Sigma_{\rm{C}}^{\Gamma_{\rm{X}}}$ presented in Fig.~\ref{fig:tcne_se}, from which we discuss the \textit{qualitative} $\Gamma_{\rm{X}}$ effect on the QP energy. We emphasize that this self-energy represents the vertex correction part to the self-energy only ($\Sigma_{\rm{C}}^{G_0W_0^{tc}\Gamma_{\rm{X}}}- \Sigma_{\rm{C}}^{G_0W_0^{tc}}$). The total self-energy curves (Fig.~\ref{fig:tcne_space}) show $\frac{\dd{\Sigma{\omega}}}{\dd{\omega}} \le 0$ at the QP energy, as indicated by the QP renormalization.\cite{fetter2003,Martin2016}. However, the individual components of the self-energy may not satisfy this inequality.

In Fig.~\ref{fig:tcne_se}a, the dashed line indicates the $\alpha$-$G_0W_0^{tc}$ QP energy, which crosses the blue curve at a negative vertical coordinate. This means the value of $\Sigma_{\rm{C}}^{\Gamma_{\rm{X}}}$ around the QP energy is negative. Given that $\beta$ is positive, the negative $\Sigma_{\rm{C}}^{\Gamma_{\rm{X}}}$ corresponds to the stabilization effect on the HOMO energy of TCNE shown in Table~\ref{tab:acceptor}. When the $\pi$ bonds are excluded from the active space, this stabilization effect disappears as the dashed line crosses the red curve at a positive vertical coordinate. Furthermore, we assume the states in the full Hilbert space can be propagated by the effective Hamiltonian with $\alpha=1$, i.e., no active space is constructed and the exchange interaction is unscreened. As indicated by the black curve in  Fig.~\ref{fig:tcne_se}a, the stabilization effect is restored. The difference among the three $\Sigma_{\rm{C}}^{\Gamma_{\rm{X}}}$ suggests that the occupied states are required in the active space to cause the stabilization effect on the QP energy.

Analogously, the crossing between the blue curve and the dashed line in Fig.~\ref{fig:tcne_se}b indicates a positive value of $\Sigma_{\rm{C}}^{\Gamma_{\rm{X}}}$, corresponding to the destabilization effect of $\Gamma_{\rm{X}}$ on the LUMO energy. However, when the $\pi^*$ bonds are excluded (the red curve) from the active space, this destabilization effect vanishes. The expansion to the full Hilbert space (equivalent to having no active space) recovers this destabilization effect (black curve in Fig.~\ref{fig:tcne_se}b). In contrast with the HOMO case, the unoccupied states are necessary in the active space for destabilizing the LUMO energy.

The observations above imply that the $\Gamma_{\rm{X}}$ effect on the HOMO energy depends mainly on the occupied (hole) states that are ``turned on'' in the active space. On the other hand, the $\Gamma_{\rm{X}}$ for the LUMO stems primarily from the unoccupied (electron) states included in the active space. This delivers an important message for the failure of simple rescaled $\Gamma_{\rm{X}}$ correction to the LUMO QP energy. The static exchange $\Sigma_{\rm{X}}$ is calculated using only the occupied (hole) states, while the LUMO state (corresponding to an injected electron) interacts mainly with the states above the Fermi level. The derived $\Gamma_{\rm{X}}$ should be dynamically screened, and its fraction cannot be trivially estimated based on the nonlocal exchange. In fact, to reproduce the experimental value of $-$2.30 eV,\cite{Lyons1976} one would need to set $\beta=0.2$, which implies a 
 substantial ``exchange reduction'' compared to its bare strength. Nevertheless, electrodynamic screening is apparently not delivering that. We surmise that the culprit is in the restriction to only density and density matrix interactions: it is likely that in small-scale systems with relatively low electron density, other types of explicit QP-QP interactions (i.e., classes of explicit two-body scatterings that are neglected here) become more important.\cite{Zaera2022} This direction will be explored in the future.

\begin{table*}
\centering
\caption{HOMO/LUMO energies and fundamental gaps of the single TCNE and TCNQ molecules computed by various methods. All energy values are in eV unit.}
\begin{threeparttable}[b]
\begin{tabular}{cccccc|ccccc}
\hline
\multirow{2}{*}{} & \multicolumn{5}{c|}{TCNE}                              & \multicolumn{5}{c}{TCNQ}                             \\
                  & DFT   & $G_0W_0$   & $\alpha$-$G_0W_0^{tc}$ & $\alpha$-$G_0W_0^{tc}\Gamma_{\beta}$   & Exp.$^1$         & DFT   & $G_0W_0$  & $\alpha$-$G_0W_0^{tc}$ & $\alpha$-$G_0W_0^{tc}\Gamma_{\beta}$   & Exp.$^1$        \\ \hline
HOMO              & -8.63 & -11.15 & -11.82 & -12.91$^2$/-13.02$^3$ & -11.79 & -7.78 & -9.63 & -9.46  & -10.70$^2$/-10.71$^3$ & -9.61 \\
LUMO              & -5.85 & -3.77  & -2.73  & -0.45$^4$         & -2.30   & -6.27 & -4.96 & -3.59  & -1.47$^4$         & -2.80 \\
Gap               & 2.76  & 7.38   & 9.09   & 10.18$^5$/10.29$^6$   & 9.49         & 1.51  & 4.66  & 5.86   & 7.11$^5$/7.12$^6$     & 6.81        \\ \hline
\end{tabular}
\begin{tablenotes}
\small
\item[1] experimental results are reported as $-$IP and $-$EA and are taken from Refs.~\citenum{Houk1976,Lyons1976,Ikemoto1974,Klots1974}
\item[2] $\Sigma_{\rm{C}}^{\Gamma_{\rm{X}}}$ term is rescaled by $\beta^0$
\item[3] $\Sigma_{\rm{C}}^{\Gamma_{\rm{X}}}$ term is rescaled by $\beta^{sc}$
\item [4] $\Sigma_{\rm{C}}^{\Gamma_{\rm{X}}}$ term is rescaled by 1
\item[5] gap is calculated by the HOMO with $\beta^{0}$ and the LUMO taken from $\alpha$-$G_0W_0^{tc}$
\item[6] gap is calculated by the HOMO with $\beta^{sc}$ and the LUMO taken from $\alpha$-$G_0W_0^{tc}$
\end{tablenotes}
\end{threeparttable}
\label{tab:acceptor}
\end{table*}

For the TCNQ molecule, 12 $\pi$ and 12 $\pi^*$ bonds constitute the active space. The representative orbitals are shown in Fig.~\ref{fig:acceptor}, and the entire set of basis is provided in Fig.~\ref{fig:acceptor_basis}. The computed rescaling factors for the occupied and unoccupied states are consistent with the TCNE case. Compared to the $G_0W_0$ result, the inclusion of $\Gamma_{W}$ ($\alpha$-$G_0W_0^{tc})$ slightly shifts the HOMO energy up, while the rescaled $\Sigma_{\rm{C}}^{\Gamma_{\rm{X}}}$ pulls the energy down by $\sim$1 eV. Similar to the TCNE case, the $G_0W_0$ approach overestimates the EA by more than 2 eV (Table~\ref{tab:acceptor}). The $\Gamma_W$ correction in $\alpha$-$G_0W_0^{tc}$ improves it by $\sim$1.4 eV, and the remaining part is attributed to the $\Gamma_{\rm{X}}$ term. However, the fraction of $\Gamma_{\rm{X}}$ is still not quantitatively determined through the simple rescaling approach. The LUMO energy becomes $-$1.47 eV with $\beta=1$, which underestimates the EA by $>$1 eV. When using $\beta=0.2$, the value ($-$3.22 eV) differs from the experiment\cite{Klots1974} ($-$2.80 eV), indicating the uniqueness of $\beta$ in each system.

Given the challenge of determining the fraction of $\Gamma_{\rm{X}}$ for the LUMO state, we do not seek to describe its effect on the LUMO QP energy quantitatively. When referring to the fundamental gaps estimated with $\alpha$-$G_0W^{tc}_0\Gamma_\beta$, we consider the HOMO value from the $\alpha$-$G_0W_0^{tc}\Gamma_\beta$ results and the LUMO energy from the $\alpha$-$G_0W_0^{tc}$ ones. Even though further exploration on the type of explicit QP-QP couplings is warranted, the fundamental gaps of the two test systems are already improved by more than 2 eV upon the $G_0W_0$ ones and agree well with the experiments, indicating the importance of the embedded vertex corrections.

\subsection{Donor-Acceptor Dimers}
\begin{figure*}
    \centering
    \includegraphics[width=0.8\textwidth]{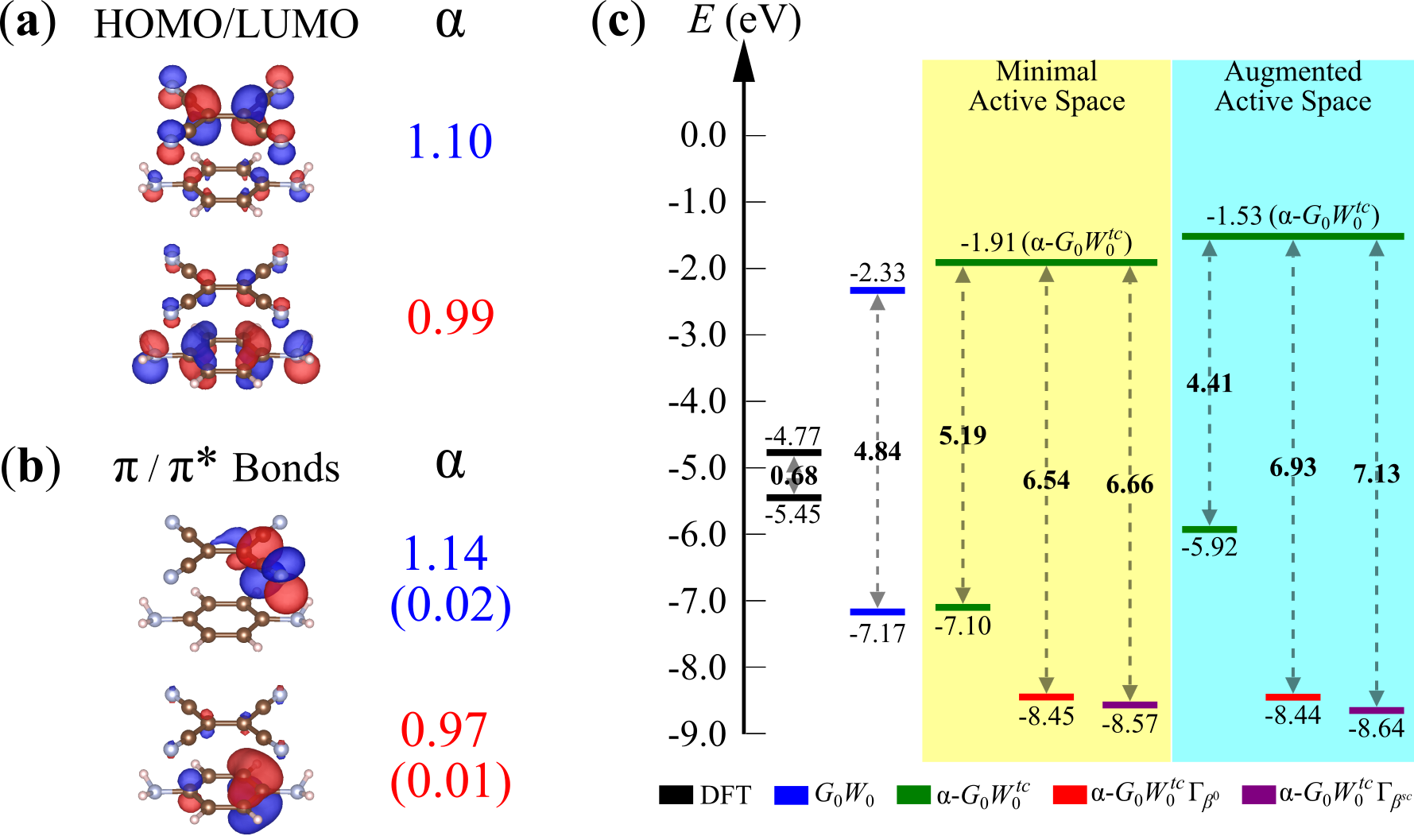}
    \caption{Computational results of the B-T dimer: (a) HOMO (lower) and LUMO (upper) states constituting the minimal active space, and their rescaling factors; (b) representative $\pi$ (lower) and $\pi^*$ (upper) bonds constituting the augmented active space, and their average rescaling factors with standard deviations in parenthesis; (c) HOMO/LUMO energies and fundamental gaps computed by various methods. The yellow and blue regions denote the two different active spaces.}
    \label{fig:BT_dimer}
\end{figure*}

In the next step, we explore systems containing two weakly bound donor-acceptor molecules, in which the active space needs to account for the charge-transfer excitation. Acene compounds and tetracyanoethylene (TCNE) derivatives have been used in computations as charge-transfer dimer models.\cite{Hayashi1997,Stein2009,Mei2020} In practice, acenes and C$_{60}$ are popular donor-acceptor materials used in electronic devices.\cite{Yoo2004,Chu2005,Yang2007,Zhang2020} This section reports computational results for two donor-acceptor combinations: 1,4-benzenediamine-TCNE (B-T) dimer and tetracene-C$_{60}$ (Tc-C) dimer. Electron and hole states participating in the charge-transfer process are chosen for the active space. The simplest choice for the charge-transfer active space is the construction from the HOMO and LUMO states, constituting the minimal charge-transfer active space. The orbitals and their rescaling factors are shown in Fig.~\ref{fig:BT_dimer}a. The nonlocal exchange is only slightly attenuated for the HOMO, while significant exchange-strengthening is observed for the LUMO state. The results are sensitive to the precise amount of exchange rescaling: a 10\% change in the nonlocal exchange interaction corresponds to about 1 eV change in energy due to the magnitude of $\Sigma_{\rm{X}}$. Therefore, the correlation contribution is especially significant for the unoccupied states.

To define an active space based on localized orbitals, we use the regional orbital localization\cite{Weng2022} on the donor and acceptor for the occupied and unoccupied subspaces. Fig.~\ref{fig:BT_dimer}b presents one $\pi$ bond on the donor and one $\pi^*$ bond on the acceptor. In total, five $\pi$ bonds  and nine $\pi^*$ bonds constitute the charge-transfer active space (Fig.~\ref{fig:bt_basis}). This new construction includes all $\pi$ components on the donor and $\pi^*$ components on the acceptor, which can be considered an ``\textit{augmented}'' active space compared to the minimal one. The $\pi$ and $\pi^*$ bonds (Fig.~\ref{fig:BT_dimer}b) are visually more localized than the canonical orbitals (Fig.~\ref{fig:BT_dimer}a), and these bonds are fundamentally equivalent in terms of their atomic-orbital components. Indeed, the factor $\alpha$ is averaged over the $\pi$ or $\pi^*$ bonds, and the standard deviations are negligible. The values are also very close to the ones in the minimal space ($<$0.05 difference).

The resulting HOMO/LUMO energies and fundamental gaps are listed in Fig.~\ref{fig:BT_dimer}c. The DFT (PBE) method, as expected, highly underestimates the gap. The $G_0W_0$ gap increases notably from 0.68 to 4.84 eV, as the HOMO and LUMO energies are shifted respectively down and up in energy. The yellow region in Fig.~\ref{fig:BT_dimer}c highlights the results using the minimal active space. Compared to $G_0W_0$, a slight energy destabilization is caused by the $\Gamma_{W}$ effect on the HOMO level. In contrast, the inclusion of $\Gamma_{\beta}$ strongly stabilizes the HOMO level by $\sim$1.4 eV. The self-consistency in $\beta$ makes a slight difference. The LUMO energy changes more than the HOMO level upon the inclusion of $\Gamma_{W}$; it is shifted up by $\sim$0.4 eV. The $\Gamma_{\rm{X}}$ contributes further destabilization (Fig.~\ref{fig:bt_dimer_eqp}b), which is consistent with our observations in the TCNE and TCNQ cases. The resulting fundamental gap from $\alpha$-$G_0W_0^{tc}\Gamma_\beta$ is $\sim$1.8 eV greater than the $G_0W_0$ one. 

When the active space is augmented (the cyan region), a larger portion of the single-particle space is treated by the effective Hamiltonian. The changes caused by $\Gamma_{W}$ to the QP energies become much more noticeable. The HOMO energy with $\Gamma_{W}$ is increased by $\sim$1.2 eV compared to $G_0W_0$. Note that when the minimal active space is used, the change is merely 0.07 eV. The following $\Gamma_{\beta}$ counters this contribution and stabilizes the HOMO energy to $-$8.44 eV (with $\beta^0$) and $-$8.64 eV (with $\beta^{sc}$). The energy change from $-$7.17 eV ($G_0W_0$) to $-$8.64 eV ($\alpha$-$G_0W_0^{tc}\Gamma_{\beta}$) agrees excellently with the self-consistent field localized orbital scaling correction (SCF-LOSC) approach\cite{Mei2020} for this dimer system. The LUMO energy with $\Gamma_{W}$ is also shifted up to $-$1.53 eV, rendering a 2-eV larger gap (the red and purple bars in Fig.~\ref{fig:BT_dimer}c) than the $G_0W_0$ one.

\begin{figure*}
    \centering
    \includegraphics[width=0.8\textwidth]{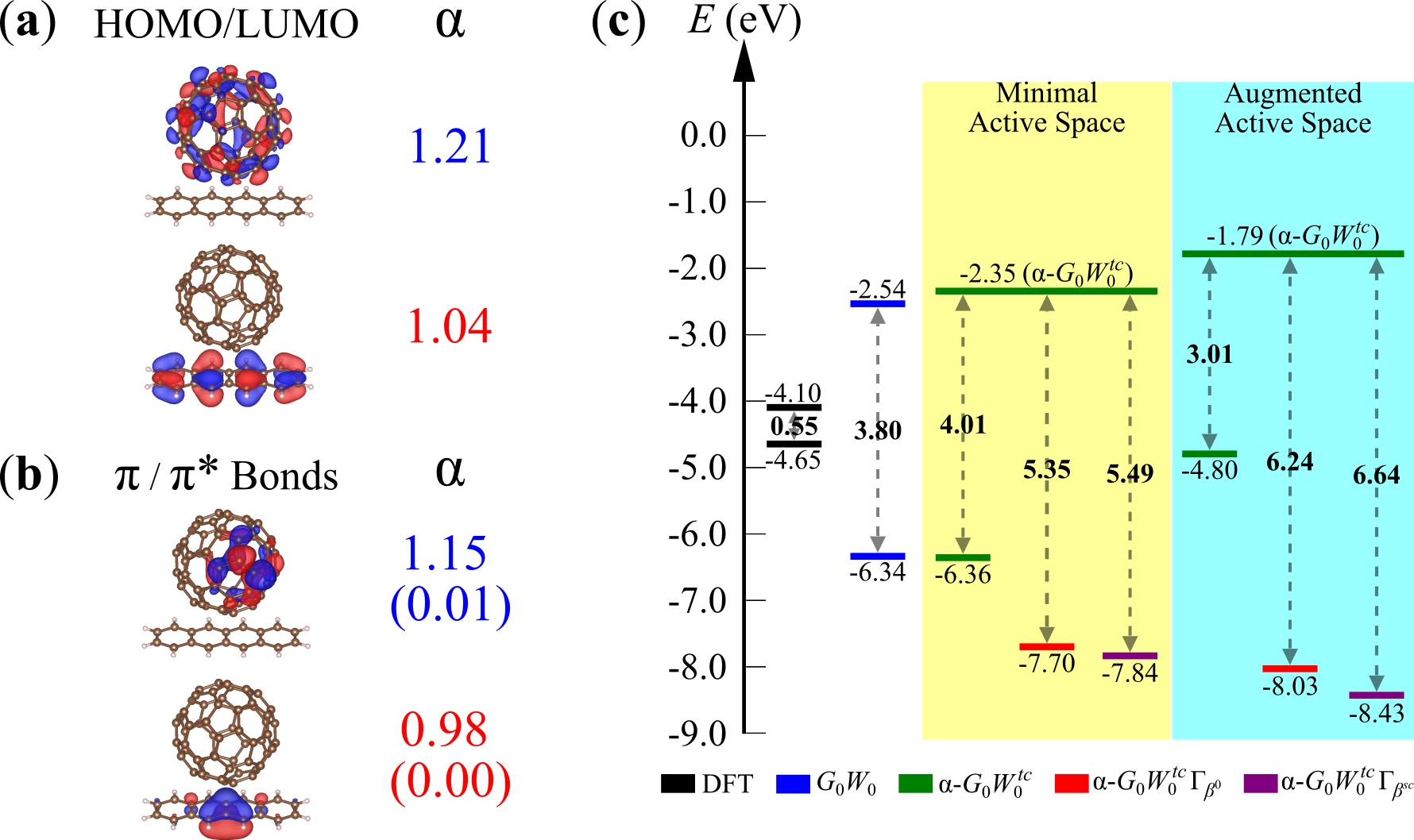}
    \caption{Computational results of the Tc-C dimer: (a) HOMO (lower) and LUMO (upper) states constituting the minimal active space, and their rescaling factors; (b) representative $\pi$ (lower) and $\pi^*$ (upper) bonds constituting the augmented active space, and their average rescaling factors with standard deviations in parenthesis; (c) HOMO/LUMO energies and fundamental gaps computed by various methods. The yellow and blue regions denote the two different active spaces.}
    \label{fig:TC_dimer}
\end{figure*}

Next, we apply the same active space construction strategy to the Tc-C dimer. Similar to the B-T dimer, the minimal active space consists of the HOMO and LUMO states (Fig.~\ref{fig:TC_dimer}a), which are localized respectively on the donor and the acceptor. The rescaling factors for these two states are both greater than 1. Unlike the previous cases, the HOMO state exhibits correlation stabilization, i.e., the QP energy decreases due to both exchange and correlation interactions. However, this picture depends strongly on the active space definition: when the space size is expanded to all $\pi$ components, a more common situation arises, in which the nonlocal exchange decreases the QP energy while the correlation counterbalances that. The factor $\alpha$ averaged over five $\pi$ bonds becomes smaller than 1 with a negligible standard deviation. For the C$_{60}$, the averaged $\alpha$ is sampled by six $\pi^*$ bonds using the geometric symmetry. The standard deviation is merely 0.01, and the average value (1.15) is thus sufficient to represent the entire set of 30 $\pi^*$ bonds (see Fig.~\ref{fig:tc_basis} in the supplemental material).

The computed energies of the Tc-C dimer are summarized in Fig.~\ref{fig:TC_dimer}c. When using the minimal active space (the yellow region) in $\alpha$-$G_0W_0^{tc}$, the HOMO energy is hardly affected by $\Gamma_{W}$, and the LUMO level is shifted up by 0.09 eV only. The $\Gamma_{\beta}$ effect is found the same as in the B-T dimer: the HOMO state is profoundly stabilized, resulting in a gap 1.5-eV larger than the $G_0W_0$ one; the self-consistency in $\beta$ mildly affects the gap by $<$0.2 eV. 

The cyan region in Fig.~\ref{fig:TC_dimer}c highlights the results using the $\pi$+$\pi^*$ active space. The Tc-C dimer behaves consistently with the B-T dimer when changing from the minimal active space to the augmented one: the HOMO energy predicted by $\alpha$-$G_0W_0^{tc}$ is significantly elevated by $\sim$1.6 eV from $G_0W_0$ due to $\Gamma_{W}$; the LUMO energy increases by $\sim$0.8 eV. The $\alpha$-$G_0W_0^{tc}$ gap (the green bar in Fig.~\ref{fig:TC_dimer}c) is thus $\sim$0.8 eV smaller than the $G_0W_0$ one. However, $\Gamma_{\beta}$ contributes oppositely to the gap by shifting the HOMO level down to $-$8.03 eV and $-$8.43 eV (with self-consistent $\beta$). The $\Gamma_{\rm{X}}$ effect on the LUMO energy behaves consistently with the previous systems (Fig.~\ref{fig:tcene_C60_eqp}b). In total, the fundamental gap of the Tc-C dimer is increased by $\sim$2.5 eV from the $G_0W_0$ one.

To conclude from the donor-acceptor dimer studies, the embedded vertex shows a clear and non-trivial dependence on the active space size, significantly affecting the fundamental gap; yet some important generalizations can be already deduced. For the HOMO state, the minimal active space implementation of $\Gamma_{W}$ makes little difference (compared to $G_0W_0$), while the augmented space notably corrects the QP energies. The $\Gamma_{\beta}$ contribution is critical to generate quantitative predictions for the HOMO QP energy. Compared to HOMO, the LUMO QP energy exhibits a more pronounced dependence on the active space size. In general, the vertex-corrected fundamental gaps are more than 2 eV larger than the $G_0W_0$ ones. Most importantly, the results of B-T dimer show a remarkable agreement between the HOMO QP energies obtained with $\alpha$-$G_0W_0^{tc}\Gamma_{\beta}$ for the two very distinct definitions of the active space. Based on this, it seems that when the vertex corrections are applied consistently to both the $W$ term and the self-energy, the dependence on the active space definition diminishes. Further investigation is necessary to clarify if this is a universal feature of the $\alpha$-$G_0W_0^{tc}\Gamma_{\beta}$ approach.

\subsection{p-FBT Polymer and Tc-C Double Layers}
The embedded vertex corrections consistently and significantly change the QP energies of the small molecules and molecular dimers investigated above. The embedding approach with two distinct active space selections is further applied to periodic systems, particularly 1D and 2D materials. Poly-fluorene-benzothiadiazole (p-FBT) is a typical donor-acceptor copolymer widely used in organic electronics.\cite{Mai2013,Mai2015,Cui2018} Charge excitation energies and fundamental gaps of these functional materials are crucial parameters in understanding their opto-electronic properties. Furthermore, we extend the Tc-C dimer in the previous section to a 2D double-layer model, which is closer to the solid-state thin-film environment and represents a minimal surface model of an organic heterojunction. For fair comparisons, The inter-layer distance takes the same value as that of the Tc-C dimer. Unlike isolated molecules, the states form band structures, and the active space is represented by energy bands. Note that for these two periodic systems, we consider only the top valence band (TVB) and bottom conduction band (BCB) for simplicity when constructing the active space. Due to our real-space implementation, the TVB or BCB is represented by a set of (more than one) orbitals, which is equivalent to the k-point sampling in the reciprocal space. We illustrate the methodology with two distinct representations: Bloch states and PM Wannier functions. The latter is obtained by unitary transforming the former. The orbital space is thus identical in dimensions, but the rescaling factors $\alpha$, in principle, differ and lead to distinct effective Hamiltonians (Eq.~\eqref{eqn:eQPH}).

\begin{figure*}
    \centering
    \includegraphics[width=0.8\textwidth]{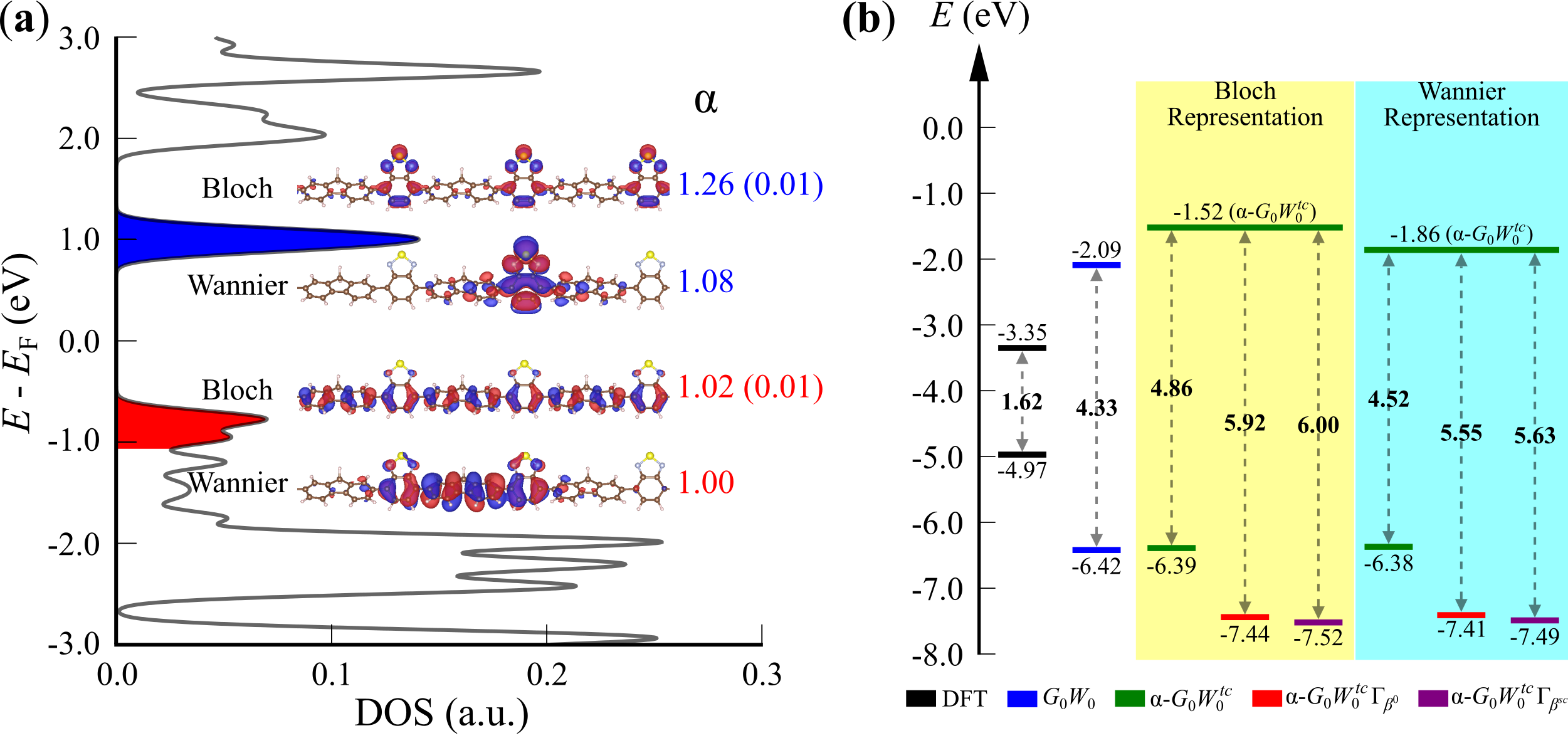}
    \caption{Computational results of the p-FBT polymer: (a) density of states with the top valence band colored in red and the bottom conduction band colored in blue. Active space are represented by either Bloch states or Wannier functions with the corresponding (averaged) rescaling factors and standard deviations in parenthesis; (b) valence band maximum/conduction band minimum energies and fundamental gaps computed by various methods. The yellow and blue regions denote the two different representations of the active space.}
    \label{fig:pFBT}
\end{figure*}

First, we exemplify the embedding scheme with the p-FBT system. Fig.~\ref{fig:pFBT}a shows a fraction of the polymer supercell, and in fact, eight repeated units are used for converging the real-space calculations. On the plot of density-of-states (Fig.~\ref{fig:pFBT}a), the highlighted regions correspond to the TVB (red) and BCB (blue) in the band structure. Due to the supercell size, eight canonical KS states are found in each band with distinct crystal momentum within the first Brillouin zone, forming the Bloch representation. Orbital localization generates eight Wannier functions. 

Typical Bloch and Wannier states are shown in Fig.~\ref{fig:pFBT}a, and the entire set of orbitals is provided in Fig.~\ref{fig:fbt_basis}. In the Bloch representation, the TVB states are delocalized along the $\pi$-conjugated backbone, while the BCB states are localized on the acceptor units. The factor $\alpha$ is averaged over eight states with negligible standard deviations (0.01). For the TVB, the $\alpha$ is slightly greater than 1, while the BCB one is much larger. These results agree with the calculations for molecules in the previous sections. In the Wannier basis, the functions are localized in each unit cell. Within the cell, the TVB Wannier function is delocalized, while the BCB one is mainly localized on the acceptor. Due to the translational symmetry, only one Wannier function is needed to represent a band when computing the factor $\alpha$. Yet, all the eight Wannier functions are needed for the active space projector (Eq.~\eqref{eqn:activep}). The $\alpha$ of Wannier representation is very close to the Bloch one for the TVB (1.00 versus 1.02) but not for the BCB (1.08 versus 1.26). In the Wannier basis, the degree of exchange rescaling is less dependent on the particular orbital type.

Fig.~\ref{fig:pFBT}b summarizes the computed results for the valence band maximum (VBM) and conduction band minimum (CBM) states of the p-FBT system. In the Bloch representation (the yellow region), the VBM remains nearly unchanged by $\Gamma_{W}$ ($\alpha$-$G_0W_0^{tc}$), while the CBM is shifted up by $\sim$0.5 eV. Consistent with the previous results, the $\Gamma_{\beta}$ term corrects the VBM energy by $\sim$1 eV to $-$7.44 eV, and the self-consistency does not make a sizable difference ($<$0.1 eV). The results in the Wannier representation (the cyan region) show explicit $\alpha$-dependence: the VBM energy does not differ too much from the Bloch one, while the CBM becomes lower in energy due to a smaller $\alpha$ being used for the unoccupied active space. The fundamental gap computed with $\alpha$-$G_0W_0^{tc}\Gamma_{\beta}$ (the red and purple bars in the cyan region of Fig.~\ref{fig:pFBT}b) is predicted to be 1.2-1.6 eV larger than the one obtained from the $G_0W_0$ calculation.

\begin{figure*}
    \centering
    \includegraphics[width=0.8\textwidth]{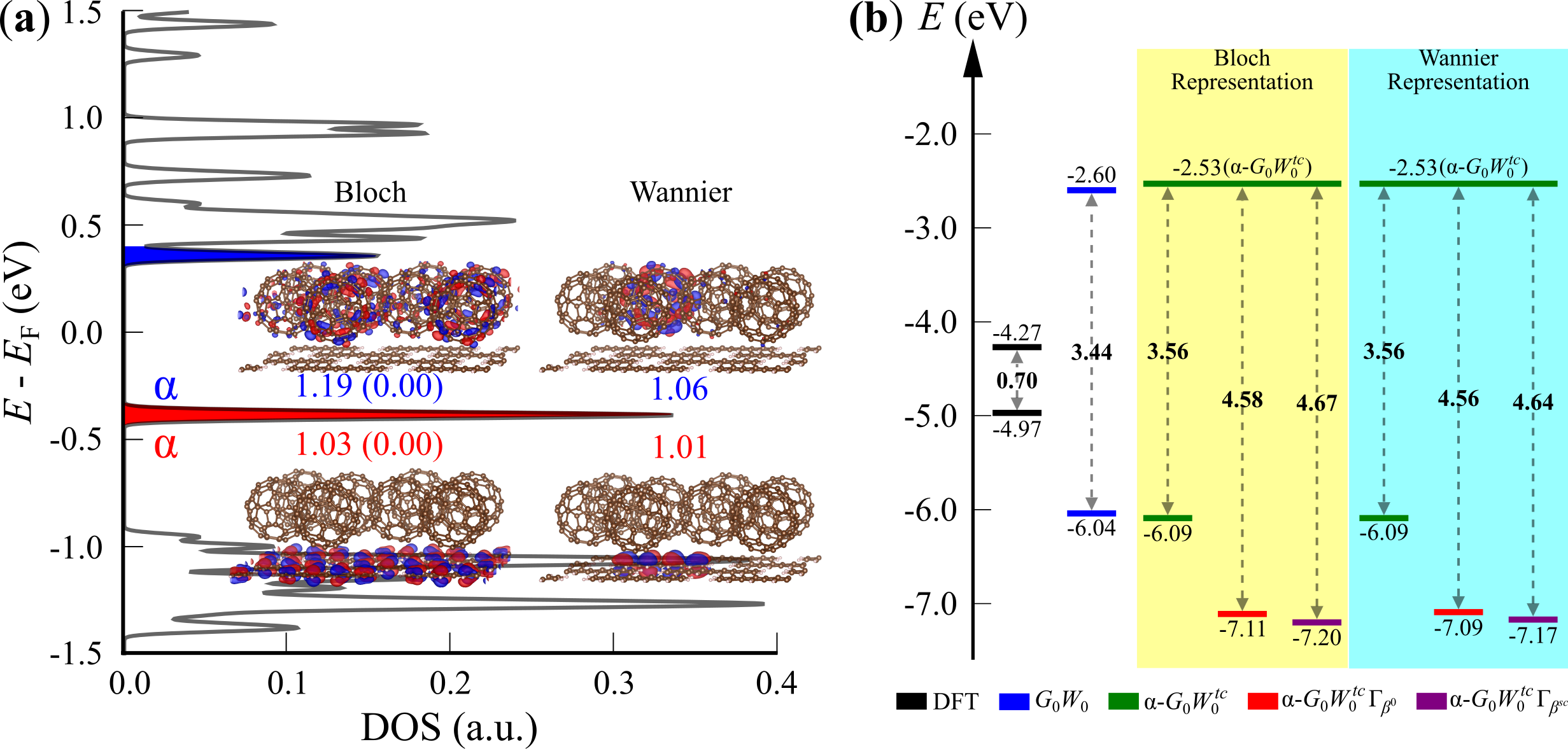}
    \caption{Computational results of Tc-C double layers: (a) density of states with the top valence band colored in red and the bottom conduction band colored in blue. Active space are represented by either Bloch states (left two states) or Wannier functions (right two states) with the corresponding (averaged) rescaling factors and standard deviations in parenthesis; (b) valence band maximum and conduction band minimum energies and fundamental gaps computed by various methods. The yellow and blue regions denote the two different representations of the active space.}
    \label{fig:TC_DL}
\end{figure*}

Similar rescaling results to the p-FBT are observed for the Tc-C double layers: the exchange interaction of the TVB is hardly rescaled in either representation; the averaged $\alpha$ of the BCB over eight Bloch states (Fig.~\ref{fig:tcdl_basis}) is greater than that using the Wannier function. However, the differences in $\alpha$ do not markedly affect the energy diagram, i.e., the computed results do not depend on the representation or the rescaling factors. We thus focus only on the yellow region in Fig.~\ref{fig:TC_DL}b. The VBM energies given by $\alpha$-$G_0W_0^{tc}$ and $G_0W_0$ are statistically the same. Moreover, the CBM of this system is also hardly affected by $\Gamma_{W}$. The $\Gamma_{\beta}$ correction then stabilizes the VBM by $\sim$1 eV, resulting in a $\sim$4.6-eV gap (the red and purple bars in Fig.~\ref{fig:TC_DL}b) that is $\sim$1.2 eV greater than the $G_0W_0$ one. The self-consistency in $\beta$ slightly shifts the energy further down. Compared with the Tc-C dimer, the solid-state renormalization of the gap is predicted to be $\sim$0.4 eV by $G_0W_0$ and $\sim$2 eV by $\alpha$-$G_0W_0^{tc}\Gamma_{\beta}$. This difference suggests that the $G_0W_0$ approach underestimates the gap renormalization in condensed systems.

The optical absorption spectrum of the tetracene-C$_{60}$ blended thin film indicates a $\sim$3-eV optical gap,\cite{Chu2005} which is similar to the $G_0W_0$ gap but deviates from the 4.6-eV result obtaine from the $\alpha$-$G_0W_0^{tc}\Gamma_{\beta}$ approach. However, the optical and fundamental gaps cannot be directly compared since several intrinsic and critical factors are not included. First of all, the exciton-binding energy significantly decreases the optical gap compared to the fundamental gap. For charge-transfer excitation in donor-acceptor systems, the exciton-binding energy follows the $1/R$ asymptotic behavior\cite{Dreuw2003,Stein2009,Baumeier2012} ($R$ is the distance between two separated charges). To estimate the exciton-binding energy, we approximate the $R$ by the distance between two respective geometric centers of the donor and acceptor layers; the bare Coulomb attraction is $\sim$2 eV between two integer charges separated by $\sim$12 Bohr. If this estimated energy is subtracted from the 4.6-eV result, the computed gap ($\sim$2.6 eV) becomes much closer to the measured optical gap. Second, the neglected electron-phonon couplings in the solid-state thin-film environment also make non-trivial contributions to the gap renormalization, in which the fundamental gap often becomes smaller.\cite{Giustino2010,Antonius2014,Antonius2015,Karsai2018} To limit the scope of this work, we do not seek to quantify the contributions above; but this rudimentary analysis demonstrates a straightforward fact that the computed fundamental gap should be significantly larger than the measured optical gap. For the double-layer system, we can compare our predictions to the scanning tunneling microscopic (STM) measurement of C$_{60}$ nanochains on bilayer pentacene.\cite{Dougherty2008} A $\sim$4-eV charge-transfer fundamental gap can be inferred from the spectra, which is also over 2 eV larger than the corresponding optical gap.\cite{Yang2007} It is thus suggestive that the embedded vertex corrections improve the fundamental gap for this tetracene-C$_{60}$ double-layer system.

\section{Conclusions and Perspective}\label{sec:candp}
This work demonstrates a real-time approach to embed vertex corrections in the $G_0W_0$ correlation self-energy. The key procedure of the proposed methodology is to perform the separation-propagation-recombination treatment on the random vectors that sample the correlation self-energy. The separation step uses an electronic active space projector to divide electron states into two components: the active part and the rest. Fundamentally, the approach is applicable to any arbitrarily chosen active space definition, as demonstrated in this work; namely, we construct the subspace with either canonical frontier orbitals (bands) or localized orbitals concerning the electron-hole pair formation. In the propagation step, the active space component is treated by a space-specific effective Hamiltonian, while the rest is treated at the RPA level. We employ a rescaled time-dependent nonlocal exchange interaction in the effective Hamiltonian, which approximates the correlation contribution and emulates the QP Hamiltonian. The use of effective Hamiltonian in time evolution introduces nonlocal vertex to the polarizability. The last step recombines the two separately time-evolved components to generate embedded vertex-corrected states that lead to correlation self-energy with vertex corrections. The effective Hamiltonian constructed in the propagation step is a crude simplification but represents a convenient and efficient way to perform the embedding. The extension of this framework is straightforward and will be explored in the near future. In the current setup, the method is successfully applied to systems ranging from small molecules to large-scale extended systems, here exemplified on 2D double layers with up to 2,500 electrons.

The vertex corrections to the polarizability (i.e., the inclusion of excitonic effects in the screening) and to the entire self-energy (including optical couplings and mitigating self-polarization errors) are discussed separately as $\Gamma_{W}$ and $\Gamma_{\beta}$. Upon the $G_0W_0$ results, $\Gamma_{W}$ is found to destabilize both the HOMO and LUMO energies in most investigated molecules. The change in QP energy exhibits strong dependence on the active space size: results from the minimal active space differ slightly from $G_0W_0$, especially for the HOMO; the destabilization effect is strongly enhanced when the active space is augmented. In the periodic systems, the frontier bands are represented by either Bloch states or Wannier functions; the latter allows an efficient evaluation of the rescaling factor for an energy band. The changes of active space representation and the corresponding exchange rescaling show contrasting influences: the CBM energy of p-FBT exhibits a clear dependence on the rescaling magnitude between two representations, while the results of the double-layer system are hardly affected.

$\Gamma_{\beta}$, in contrast, enters the self-energy and is also derived from the rescaled nonlocal exchange approximation. An appealing advantage of the scheme proposed here is the possibility to update the rescaling factor in $\Gamma_{\beta}$ in a self-consistent post-processing step, which incurs no increase of the computational cost. In general, the inclusion of $\Gamma_{\beta}$ stabilizes the HOMO energy but destabilizes the LUMO level. For the former, the $\Gamma_{\beta}$ correction reaches a quantitative agreement with the previous work on the B-T dimer.\cite{Mei2020} The self-consistency in $\beta$ leads to additional stabilization of the QP energy. 

For the LUMO state, the simplistic approach based on the rescaled nonlocal exchange interaction fails to reach a quantitative description. Further investigations on the components of the active space imply that the unoccupied (electron) states play a dominating role in the $\Sigma_{\rm{C}}^{\Gamma_{\rm{X}}}$ self-energy for the LUMO. However, the rescaling approximation involves only occupied (hole) states when calculating the exchange interaction, which is the surmised culprit. Further exploration is needed to approximate the correlation contribution for deriving the vertex function. Generally, the fundamental gaps of the investigated systems are increased by 1-3 eV. This improvement is especially significant in our donor-acceptor double-layer system, as the results agree excellently with the related experiments.

Finally, we note that if the vertex corrections are used internally and consistently in both the $W$ term and the self-energy, the results exhibit a much weaker dependence on the choice of the active space. Ref.\citenum{Zaera2022} points out that the vertex corrections arise naturally as a consequence of functional self-consistency, in which new classes of interactions (embodies by the vertex term) appear by evaluating the functional derivatives of the self-energy in \textit{both} terms, i.e., $W$ and $\Sigma_{\rm{C}}$. This observation warrants further exploration to determine if indeed the inclusion of both $\Gamma_W$ and $\Gamma_{\beta}$ help to mitigate the active space selection conundrum.

In summary, our proposed methodology combines the concepts of stochastic sampling, Pipek-Mezey localized orbitals, space-specific Hamiltonian, and separation-propagation-recombination treatment. It provides an efficient and direct way to include nonlocal vertex corrections to the $GW$ self-energy. We believe that this work on embedding methods within the Green's function framework will stimulate more attempts to handle vertex corrections in large condensed matter systems.

\section*{Supplemental Material}
The supplemental material provides the details of theory and computation, as well as supplementary tables and figures indicated in the texts.

\begin{acknowledgements}
The authors want to acknowledge Prof.~Thuc-Quyen Nguyen for useful discussions on the donor and acceptor molecules. This work was supported by the NSF CAREER award through Grant No. DMR-1945098. The calculations were performed as part of the XSEDE computational Project No. TG-CHE180051. Use was made of computational facilities purchased with funds from the National Science Foundation (CNS-1725797) and administered by the Center for Scientific Computing (CSC). The CSC is supported by the California NanoSystems Institute and the Materials Research Science and Engineering Center (MRSEC; NSF DMR 1720256) at UC Santa Barbara.
\end{acknowledgements}

\section*{DATA AVAILABILITY}
The data that support the findings of this study are available from the corresponding author upon reasonable request.

\bibliography{embedded_vertex}

\end{document}